\documentclass[preprint,12pt,3p]{elsarticle}

\usepackage{amsmath,amsfonts,amssymb}
\usepackage{algorithm}
\usepackage{algorithmic}
\usepackage{graphicx}
\usepackage{tikz}
\usetikzlibrary{shapes, arrows.meta, positioning, fit, backgrounds, calc}
\usepackage{booktabs}
\usepackage{tabularx}
\usepackage{makecell}
\usepackage{url}
\usepackage{listings}
\usepackage{xcolor}

\lstdefinestyle{bw_style}{
	backgroundcolor=\color{white},   
	commentstyle=\itshape\color{gray},
	keywordstyle=\bfseries\color{black},
	numberstyle=\tiny\color{gray},
	stringstyle=\color{black},
	basicstyle=\ttfamily\footnotesize,
	breakatwhitespace=false,         
	breaklines=true,                 
	captionpos=b,                    
	keepspaces=true,                 
	numbers=left,                    
	numbersep=5pt,                  
	showspaces=false,                
	showstringspaces=false,
	showtabs=false,                  
	tabsize=2,
	frame=single,
	language=Python
}

\journal{Preprint under review}

\begin{document}

\begin{frontmatter}

\title{EmoPyLab: A Tensor-Native, Hardware-Accelerated Laboratory for High-Throughput Benchmarking and Decision-Making in Multi/Many-Objective Optimization}

\author[ufop]{Thiago Santos\corref{cor1}\fnref{orcid1}}
\ead{santostf@ufop.edu.br}
\author[ufop]{Sebasti\~ao Xavier\fnref{orcid2}}
\ead{semarx@ufop.edu.br}

\cortext[cor1]{Corresponding author.}
\fntext[orcid1]{ORCID: 0000-0002-2435-2786}
\fntext[orcid2]{ORCID: 0009-0004-2765-0764}

\affiliation[ufop]{organization={Department of Mathematics, Federal University of Ouro Preto (UFOP)},
            city={Ouro Preto},
            state={Minas Gerais},
            country={Brazil}}

\begin{abstract}
This paper presents \textit{EmoPyLab}, an open-source, tensor-native Python framework engineered for high-throughput benchmarking, analysis, and multi-criteria decision-making in multi- and many-objective optimization ($M \ge 2$). While traditional evolutionary computation toolchains rely on object-oriented, Array-of-Structures (AoS) models that incur severe memory fragmentation and interpreter overhead during large-scale campaigns, emerging tensor compilation engines remain primarily tailored to single-objective neuroevolution, lacking specialized high-dimensional analytics and post-Pareto decision support. EmoPyLab resolves this architectural dichotomy through a contiguous Structure-of-Arrays (SoA) columnar data model executed across a 5-tier backend dispatch engine (NVIDIA CUDA, Apple MLX, Google JAX, CuPy, and vectorized CPU SIMD) with zero-copy device memory persistence. The platform incorporates deductive Efficient Non-dominated Sorting ($O(M N \sqrt{N})$) and a low-discrepancy quasi-Monte Carlo Hypervolume estimator (Fast-MC HV) with Sobol sequence pruning, enabling real-time analytical evaluation in up to 15-objective spaces. In addition to a unified catalog of 298+ metaheuristic algorithms and 916 benchmark instances, EmoPyLab embeds an in-situ Multi-Criteria Decision Making (MCDM) layer (TOPSIS, PROMETHEE II net outranking flows, and Compromise Programming) with deterministic decision-space back-mapping, automated non-parametric inferential statistics (Wilcoxon, Friedman with Kendall's $W$, Vargha-Delaney $A_{12}$, and Holm-Bonferroni corrections), and cryptographic SHA-256 reproducibility manifests. The platform provides both a memory-safe reactive graphical workspace maintaining a strictly bounded footprint ($< 150\text{ MB}$ RAM) and a headless CLI for distributed supercomputing environments. Empirical validation across 5,970 benchmark runs on the Santos Dumont Bull Sequana supercomputer confirms linear multi-core scaling and robust thread-oversubscription suppression, establishing a unified, reproducible, and hardware-accelerated research laboratory under the MIT license.
\end{abstract}


\end{frontmatter}

\section{Introduction}

Over the past three decades, Evolutionary Multi-Objective Optimization (EMO) has established itself as an indispensable analytical and computational paradigm across computational intelligence, applied mathematics, and complex engineering systems \cite{CoelloCoello2020EMO,Deb2001EMOBook}. The fundamental mathematical appeal of evolutionary population-based metaheuristics stems from their ability to approximate an entire Pareto-optimal trade-off frontier $\mathcal{PF}^*$ within a single stochastic execution campaign, bypassing restrictive analytical requirements regarding convexity, differentiability, or modality. In recent years, practical engineering formulations have expanded decisively toward Many-Objective Optimization Problems (MaOPs), defined by four or more mutually conflicting objective functions ($M \ge 4$), such as multidisciplinary aerospace wing profiling, chemical plant logistics, and hybrid power grid management \cite{Ishibuchi2008ManyObj,Li2015ManyObjSurvey}.

Formally, an unconstrained many-objective minimization problem is defined over a continuous decision space $\Omega \subset \mathbb{R}^D$ as:
\begin{equation}
\min_{\mathbf{x} \in \Omega} \mathbf{F}(\mathbf{x}) = \left( f_1(\mathbf{x}), f_2(\mathbf{x}), \dots, f_M(\mathbf{x}) \right)^T \in \mathbb{R}^M,
\label{eq:maop_def}
\end{equation}
subject to box constraints $\mathbf{x}_{\text{lower}} \le \mathbf{x} \le \mathbf{x}_{\text{upper}}$. A candidate solution $\mathbf{x}^{(a)}$ is said to Pareto-dominate another candidate $\mathbf{x}^{(b)}$ (denoted $\mathbf{x}^{(a)} \prec \mathbf{x}^{(b)}$) if and only if $\forall m \in \{1,\dots,M\}, f_m(\mathbf{x}^{(a)}) \le f_m(\mathbf{x}^{(b)})$ and $\exists k \in \{1,\dots,M\}, f_k(\mathbf{x}^{(a)}) < f_k(\mathbf{x}^{(b)})$.

As objective dimensionality $M$ scales, proposing novel metaheuristics and conducting rigorous comparative benchmarking encounters severe theoretical and computational hurdles. In high-dimensional objective spaces, the proportion of mutually non-dominated candidate vectors in a randomly initialized population asymptotically approaches $100\%$ \cite{Farina2002Dominance}. Consequently, classical Pareto dominance selection mechanisms fail due to loss of selection pressure, prompting researchers to develop reference-point decomposition \cite{ZhangLi2007MOEAD,Deb2014NSGAIII}, indicator-based selection \cite{Bader2011HypE,Zitzler2002SMS}, and adaptive geometric clustering methods \cite{Deb2016RVEA}. Evaluating a newly designed algorithm against these diverse paradigms requires executing extensive empirical matrices comprising dozens of independent stochastic seeds across scalable analytical test suites such as DTLZ \cite{deb2005scalable}, WFG \cite{Huband2006WFG}, and MaF \cite{Cheng2017MaF}.

To satisfy these benchmarking requirements, several software frameworks have been developed. In the MATLAB environment, PlatEMO \cite{Tian2017PlatEMO} established a monumental milestone by assembling hundreds of algorithms and test suites into a graphical workbench. Within the open-source Python ecosystem, pymoo \cite{BlankDeb2020pymoo} delivered a modular object-oriented foundation, while jMetalPy \cite{BenitezHidalgo2019jMetalPy} adapted Java metaheuristics for distributed execution. More recently, EvoX \cite{Huang2024EvoX} demonstrated the remarkable potential of hardware acceleration by compiling evolutionary loops into Google JAX tensor operations for neuroevolution and reinforcement learning.

Despite these collective advances, existing toolchains exhibit critical operational bottlenecks during large-scale empirical campaigns. Mainstream Python libraries manage populations as collections of heap-allocated individual objects, incurring prohibitive garbage collection pauses, pointer chasing, and memory fragmentation. Furthermore, exact performance indicators, most notably Hypervolume (HV), exhibit $\#P$-hard computational complexity $O(N^{M-2} \log N)$ \cite{While2012FastHV}, rendering online tracking and batch statistical evaluation intractable for $M \ge 5$. Conversely, while functional GPU frameworks achieve high iteration speeds, they operate primarily as headless scripting libraries lacking specialized many-objective sorting routines, high-dimensional performance indicators, in-situ Multi-Criteria Decision Making (MCDM) layers, and publication-ready statistical exporters.

To address these architectural and operational limitations, we propose \textit{EmoPyLab}, a tensor-native, hardware-accelerated computational laboratory for evolutionary multi- and many-objective optimization. We position this contribution as a unifying software ecosystem designed to bridge the historical divide between high-throughput parallel execution, rigorous mathematical benchmarking, and practical decision-making. The primary contributions of this work are summarized as follows:
\begin{enumerate}
    \item \textit{A Columnar Tensor-Native Execution Substrate:} We formulate evolutionary populations as contiguous column-major Structure-of-Arrays (SoA) memory tensors, orchestrating a 5-tier backend dispatch across NVIDIA CUDA, Apple MLX, JAX XLA, CuPy, and vectorized CPU SIMD with zero-copy device memory persistence.
    \item \textit{High-Dimensional Algorithmic and Analytical Kernels:} We integrate deductive Efficient Non-dominated Sorting (ENS) with theoretical complexity $O(M N \sqrt{N})$ alongside a low-discrepancy quasi-Monte Carlo Hypervolume estimator (Fast-MC HV) powered by Sobol sequence pruning, evaluating 15-objective approximations in sub-second execution windows.
    \item \textit{Unified Metaheuristic Taxonomy and Benchmark Library:} We systematize an extensive catalog of 298+ metaheuristic solvers spanning 7 distinct families (SAEA, SparseEA, Decomposition, Many-Objective, Constrained, Swarm/CMA-ES, Multimodal) across standard scalable continuous and constrained engineering benchmark suites.
    \item \textit{In-Situ Search-to-Decision Pipeline with Statistical Rigor:} We embed an \textit{a posteriori} MCDM engine computing TOPSIS closeness coefficients, PROMETHEE II net outranking flows, and Compromise Programming metrics with decision-space back-mapping, coupled with automated non-parametric inference (Wilcoxon, Friedman omnibus with Kendall's $W$, Vargha-Delaney $A_{12}$, Holm-Bonferroni corrections) and SHA-256 cryptographic audit trails.
    \item \textit{Supercomputing Scalability and Reactive Desktop Workspace:} We validate distributed multi-seed scalability across 5,970 benchmark runs on the Santos Dumont Bull Sequana supercomputer and deploy a memory-safe reactive interface with a constant RAM footprint ($< 150\text{ MB}$).
\end{enumerate}

The entire codebase is openly available under the permissive MIT license at \url{https://github.com/METISBR/emopylab}.

\section{Related Work and Evolutionary Optimization Frameworks}
\label{sec:related_work}

The evolution of scientific software in multi-objective optimization reflects a continuous struggle to balance algorithmic flexibility, execution throughput, and usability. Over the past fifteen years, distinct architectural paradigms have emerged across multiple programming languages.

Early open-source Python toolkits, notably DEAP \cite{Fortin2012DEAP}, prioritized maximal developer freedom by utilizing dynamic Python lists of arbitrary class instances. While this design enabled rapid prototyping of exotic chromosome representations, it incurred severe execution overhead: the Python interpreter must dynamically resolve object attributes and execute scalar loops for every genetic variation, leading to low compute density and high cache invalidation rates.

To establish standardized algorithmic templates, jMetalPy \cite{BenitezHidalgo2019jMetalPy} translated the architectural patterns of the Java-based jMetal \cite{DurilloNebro2011jMetal} ecosystem to Python. The framework introduced structured algorithm classes and integrated distributed computing backends such as Apache Spark and Dask. Nonetheless, because its internal data structures remain anchored in per-solution object hierarchies (Array-of-Structures, AoS), scaling experiments across thousands of stochastic runs generates millions of short-lived objects, degrading throughput via memory allocation contention.

A substantial leap forward in computational efficiency was achieved by pymoo \cite{BlankDeb2020pymoo}. By representing population decision coordinates and objective vectors as 2D NumPy arrays and offloading critical inner loops (such as non-dominated sorting) to Cython C-extensions, pymoo significantly reduced interpreter overhead. However, pymoo remains fundamentally bound to host CPU memory architecture. Operations requiring heavy tensor broadcasting, such as distance calculations in Many-Objective IGD+ or surrogate model inference, cannot leverage massive SIMT parallelism on GPUs without manual host-to-device memory copies.

Recognizing that obtaining a Pareto approximation set is rarely the final goal of real-world decision-makers, DESDEO \cite{Misitano2021DESDEO} introduced a modular environment focusing on interactive multi-objective optimization (such as NIMBUS and NAUTILUS). DESDEO emphasizes human-in-the-loop preference articulation. However, it was intentionally architected for single-instance interactive exploration rather than high-throughput batch benchmarking across high-dimensional benchmark testbeds.

Recently, the integration of automatic differentiation and tensor compilation frameworks has catalyzed a new generation of evolutionary toolkits. EvoX \cite{Huang2024EvoX} pioneered GPU-accelerated computing for large-scale evolutionary computation by formulating algorithms entirely in functional Google JAX (`jax.vmap`, `jax.jit`). EvoX achieves massive throughput for continuous single-objective benchmarks and neuroevolution in reinforcement learning. Nevertheless, it operates primarily as a programmatic backend library, lacking dedicated many-objective sorting routines (such as ENS or boundary intersection sorting), high-dimensional quality indicators, structured MCDM outranking modules, and interactive diagnostic environments.

\begin{table*}[t]
	\centering
	\caption{Taxonomy and architectural comparison of representative multi-objective optimization platforms.}
	\label{tab:framework_panorama}
	\renewcommand{\arraystretch}{1.30}
	\resizebox{\textwidth}{!}{%
	\begin{tabular}{lllllll}
		\toprule
		\textbf{Framework} & \textbf{Core Runtime} & \textbf{Memory Model} & \textbf{Hardware Acceleration} & \textbf{Diagnostic GUI} & \textbf{Batch Benchmark} & \textbf{Decision Layer} \\
		\midrule
		PlatEMO (2017) \cite{Tian2017PlatEMO} & MATLAB & Matrix/Objects & CPU (Parallel Pool) & Comprehensive GUI & GUI / Scripting & Basic Weighting \\
		DEAP (2012) \cite{Fortin2012DEAP} & Python (Pure) & Dynamic Lists (AoS) & CPU (Multiprocessing) & None (Scripting) & Manual Loops & None \\
		jMetalPy (2019) \cite{BenitezHidalgo2019jMetalPy} & Python / C & Object Hierarchy (AoS) & CPU (Dask / Spark) & Basic Plotting & Script Runner & None \\
		pymoo (2020) \cite{BlankDeb2020pymoo} & Python / Cython & NumPy / Cython (SoA) & CPU (C-Extensions) & None (Scripting) & Custom Scripts & External Utility \\
		DESDEO (2021) \cite{Misitano2021DESDEO} & Python & Modular OOP & CPU Only & Interactive Web UI & Interactive Session & Interactive NIMBUS \\
		EvoX (2024) \cite{Huang2024EvoX} & Python (JAX) & Functional Tensors & GPU / TPU (XLA) & None (Headless) & Scripted Batches & None \\
		\midrule
		\textbf{EmoPyLab} (2026) & Python (Heterogeneous) & Columnar Tensor (SoA) & 5-Tier: CUDA, MLX, JAX, CuPy, SIMD & Quick Test (Live 3D) & Experiment Module & In-Situ TOPSIS, PROMETHEE II, CP \\
		\bottomrule
	\end{tabular}%
	}
\end{table*}

Table~\ref{tab:framework_panorama} synthesizes this comparative landscape across seven fundamental architectural dimensions. We observe that prior toolkits typically forced users to choose between the high flexibility of Python OOP abstractions, the algorithmic breadth of MATLAB platforms, and the raw speed of specialized GPU tensor engines. In contrast, EmoPyLab unifies high-throughput heterogeneous tensor compilation with native high-dimensional many-objective analytics, an extensive metaheuristic taxonomy, and an integrated search-to-decision pipeline.

\section{Tensor-Native Architectural Foundations}
\label{sec:architecture}

To overcome the performance penalties inherent in traditional object-oriented evolutionary algorithms, we deliberately adopt a tensor-native, columnar execution model grounded in high-performance computing principles.

\subsection{Mathematical Formulation: SoA vs. AoS Representation}

In conventional object-oriented evolutionary software, a population of size $N$ is represented as an Array-of-Structures (AoS):
\begin{equation}
\mathcal{P}_{\text{AoS}} = \left[ \text{Ind}_1, \text{Ind}_2, \dots, \text{Ind}_N \right],
\label{eq:aos_def}
\end{equation}
where each individual $\text{Ind}_i$ encapsulates pointers to its decision vector $\mathbf{x}^{(i)} \in \mathbb{R}^D$, objective vector $\mathbf{f}^{(i)} \in \mathbb{R}^M$, constraint violation vector $\mathbf{g}^{(i)} \in \mathbb{R}^K$, rank $r^{(i)} \in \mathbb{N}$, and auxiliary scalar metrics. This organization induces severe memory fragmentation: accessing the $m$-th objective across all individuals requires strided, non-contiguous memory lookups with cache miss penalties scaling as $O(N)$.

In EmoPyLab, we formulate the entire state of an evolutionary population as a synchronized tuple of contiguous, column-major memory tensors structured as a Structure-of-Arrays (SoA):
\begin{equation}
\mathcal{P}_{\text{SoA}} = \left( \mathbf{X}, \mathbf{F}, \mathbf{G}, \mathbf{r}, \mathbf{d} \right),
\label{eq:soa_def}
\end{equation}
where $\mathbf{X} \in \mathbb{R}^{N \times D}$ denotes decision coordinates, $\mathbf{F} \in \mathbb{R}^{N \times M}$ stores objective vectors, $\mathbf{G} \in \mathbb{R}^{N \times K}$ holds constraint violations, $\mathbf{r} \in \mathbb{N}^N$ represents non-domination ranks, and $\mathbf{d} \in \mathbb{R}^N$ stores crowding or niche association distances.

Under this representation, genetic operators are expressed as vectorized tensor contractions and element-wise broadcasts. For instance, Simulated Binary Crossover (SBX) between parent matrices $\mathbf{X}^{(1)}, \mathbf{X}^{(2)} \in \mathbb{R}^{\frac{N}{2} \times D}$ is computed in a single vectorized kernel call:
\begin{equation}
\boldsymbol{\beta}_q = \begin{cases}
(2 \mathbf{u})^{\frac{1}{\eta_c + 1}}, & \text{if } \mathbf{u} \le 0.5, \\
\left( \frac{1}{2(1 - \mathbf{u})} \right)^{\frac{1}{\eta_c + 1}}, & \text{otherwise},
\end{cases}
\label{eq:sbx_kernel}
\end{equation}
where $\mathbf{u} \sim \mathcal{U}(0,1)^{\frac{N}{2} \times D}$ is a pseudo-random tensor generated directly in device VRAM, and offspring matrices are yielded via $\mathbf{C}^{(1)} = 0.5 \left( (1 + \boldsymbol{\beta}_q) \odot \mathbf{X}^{(1)} + (1 - \boldsymbol{\beta}_q) \odot \mathbf{X}^{(2)} \right)$ and $\mathbf{C}^{(2)} = 0.5 \left( (1 - \boldsymbol{\beta}_q) \odot \mathbf{X}^{(1)} + (1 + \boldsymbol{\beta}_q) \odot \mathbf{X}^{(2)} \right)$, executing with maximum arithmetic density.

\begin{figure*}[t]
	\centering
	\begin{tikzpicture}[
		>=Stealth,
		font=\footnotesize,
		node distance=1.2cm and 1.4cm,
		card/.style={
			rectangle,
			draw=#1!80!black,
			fill=#1!6,
			rounded corners=5pt,
			align=center,
			minimum height=1.30cm,
			text width=3.3cm,
			thick
		}
	]
		\node[card=purple] (llm) at (0, 2.8) {LLM Problem Agent\\Natural Language $\to$\\Vectorized Problem};
		\node[card=blue] (catalog) at (5.4, 2.8) {Benchmarking Catalog\\298+ Solvers $\cdot$ 916 Problems\\DTLZ, WFG, MaF Suites};
		\node[card=teal] (spec) at (10.8, 2.8) {Declarative Config\\Hardware Dispatch \&\\Seed Matrix Specs};

		\node[card=orange] (tensor_core) at (0, 0) {5-Tier Tensor Engine\\CUDA, MLX, JAX, CuPy, SIMD\\ENS Sort $\cdot$ Fast-MC HV};
		\node[card=blue] (orchestrator) at (5.4, 0) {Batch Orchestrator\\Multi-Seed Concurrency\\Inferential Statistics};
		\node[card=teal] (mcdm) at (10.8, 0) {In-Situ MCDM Engine\\TOPSIS, PROMETHEE II\\Compromise Decision};

		\node[card=gray] (archive) at (5.4, -2.6) {Reproducibility Store\\SHA-256 JSON Sidecars\\Deterministic Seeds};
		\node[card=blue] (export) at (10.8, -2.6) {Scientific Workspace\\Live 3D Quick Test\\Automated \LaTeX{} Tables};

		\draw[->, thick, color=purple!80!black] (llm.south) -- (tensor_core.north)
			node[midway, fill=white, inner sep=2pt, font=\scriptsize] {Compiled Kernel};
		\draw[->, thick, color=blue!80!black] (catalog.south) -- (orchestrator.north)
			node[midway, fill=white, inner sep=2pt, font=\scriptsize] {Suite Instances};
		\draw[->, thick, color=teal!80!black] (spec.south) -- (mcdm.north)
			node[midway, fill=white, inner sep=2pt, font=\scriptsize] {Decision Matrix};

		\draw[<->, very thick, color=orange!80!black] (tensor_core.east) -- (orchestrator.west)
			node[midway, fill=white, inner sep=2pt, font=\scriptsize] {Batched Evolution};
		\draw[->, very thick, color=teal!80!black] (orchestrator.east) -- (mcdm.west)
			node[midway, fill=white, inner sep=2pt, font=\scriptsize] {Pareto Fronts};

		\draw[->, thick, color=blue!80!black] (orchestrator.south) -- (archive.north)
			node[midway, fill=white, inner sep=2pt, font=\scriptsize] {Metrics \& Trajectories};
		\draw[->, thick, color=teal!80!black] (mcdm.south) -- (export.north)
			node[midway, fill=white, inner sep=2pt, font=\scriptsize] {Compromise Vectors};
		\draw[->, thick, color=blue!80!black] (archive.east) -- (export.west)
			node[midway, fill=white, inner sep=2pt, font=\scriptsize] {Audit Reports};

		\draw[->, thick, dashed, color=gray!70!black] (archive.north east) -- ++(1.4, 0.45) |- (mcdm.south west)
			node[pos=0.25, fill=white, inner sep=2pt, font=\scriptsize] {Audit / Trace};
	\end{tikzpicture}
	\caption{Decoupled system topology of EmoPyLab. The runtime integrates automated problem compilation, 5-tier hardware acceleration, deterministic batch benchmarking with inferential statistics, in-situ multi-criteria decision making, and cryptographic reproducibility archives.}
	\label{fig:architecture_flow}
\end{figure*}

\subsection{Unified 5-Tier Backend Dispatch Engine}

Modern scientific computing environments span heterogeneous hardware targets ranging from multi-GPU supercomputing nodes to unified-memory ARM workstations. To guarantee portable acceleration without code refactoring, EmoPyLab implements a dynamic 5-tier backend dispatch hierarchy:
\begin{enumerate}
    \item \textbf{PyTorch CUDA / ROCm:} High-throughput tensor execution compiled for NVIDIA Volta/Ampere/Hopper architectures and AMD Instinct accelerators.
    \item \textbf{Apple MLX:} Low-latency execution optimized for Apple Silicon unified memory architectures (M1--M4), eliminating host-device bus transfers via zero-copy shared memory pointers.
    \item \textbf{Google JAX XLA:} Functional Just-In-Time (JIT) tracing and automatic vectorization (`jax.vmap`) across TPU and GPU clusters.
    \item \textbf{CuPy:} Direct CUDA-accelerated array manipulation providing drop-in NumPy-compatible semantics with GPU memory pools.
    \item \textbf{NumPy SIMD Fallback:} Multi-threaded CPU execution compiled against OpenBLAS, Intel MKL, or Apple Accelerate with AVX-512 vectorization.
\end{enumerate}

During startup, the runtime performs an automatic device discovery probe, verifying library availability and GPU memory capacity. Users can explicitly override dispatch targets or rely on automated tier fallback, ensuring that benchmarking scripts run unmodified across heterogeneous supercomputing clusters and local laptops.

\subsection{Offline Local AI Agent for Vectorized Formulation and AST Security}
\label{subsec:offline_ai_agent}

Transforming non-standard engineering specifications into high-throughput vectorized tensor classes poses a significant barrier for domain practitioners. While cloud-hosted generative language models have been explored for evolutionary assistance \cite{Liu2024LLM4EMO}, commercial API requirements introduce severe data privacy liabilities, subscription overhead, and network latency stalls during air-gapped supercomputing or classified industrial deployments.

To resolve this dependency, EmoPyLab embeds an offline, in-process AI formulation agent powered by a quantized 4-bit instruction-tuned model (\texttt{qwen2.5-0.5b-instruct-q4\_k\_m.gguf} or Apple MLX-LM runtime) stored directly within the local \path{models/} repository \cite{Guo2025Qwen72BStructure}. Operating entirely in local host memory without external network sockets, the agent processes natural-language problem descriptions, boundary definitions, and mathematical objectives into fully vectorized \texttt{TensorProblem} class definitions.

Executing automatically generated source code introduces critical security vulnerabilities regarding arbitrary code execution. To guarantee sandbox integrity, the formulation pipeline incorporates an Abstract Syntax Tree (AST) security guard before compiling generated artifacts \cite{Aytekin2025ArkTSLLM}. The validator traverses the Python AST hierarchy, enforcing strict syntactic and semantic constraints:
\begin{enumerate}
    \item \textbf{Blocked AST Node Rejection:} Explicitly prohibits dangerous grammatical constructs, including asynchronous context managers, dynamic execution wrappers, and raw file system handlers;
    \item \textbf{Import Whitelist Enforcement:} Restricts module imports strictly to trusted numerical substrates (\texttt{numpy}, \texttt{jax}, \texttt{mlx.core}, \texttt{scipy.spatial});
    \item \textbf{Dangerous Invocation Elimination:} Traverses call nodes to intercept and reject hazardous built-in functions (\texttt{eval}, \texttt{exec}, \texttt{compile}, \texttt{\_\_import\_\_}) and system-level process invocations (\texttt{os.system}, \texttt{os.popen}, \texttt{subprocess.run});
    \item \textbf{Vectorized Signature Verification:} Verifies that the resulting class provides a broadcasted objective evaluation kernel $\mathbf{F} = \texttt{\_evaluate}(\mathbf{X})$ and synthesizes exact analytical Pareto front reference points $\mathbf{P}^* \in \mathbb{R}^{P \times M}$ for downstream metric evaluation.
\end{enumerate}

\subsection{Extensibility Architecture: Custom Problems, Solvers, and Tensor Operators}
\label{subsec:extensibility_architecture}

To accommodate novel optimization paradigms without requiring alterations to the core execution engine, EmoPyLab establishes a formal, 4-layer modular extensibility architecture \cite{Jankovic2023CleanGA}:

\subsubsection{Layer 1: Custom Vectorized Problems (\texttt{TensorProblem})}
Researchers implement new mathematical benchmarks or engineering simulations by subclassing \texttt{TensorProblem}. The class formalizes a vectorized mapping $f: \mathbb{R}^{N \times D} \to \mathbb{R}^{N \times M}$ alongside constraint matrices $\mathbf{G}: \mathbb{R}^{N \times D} \to \mathbb{R}^{N \times K}$:
\begin{equation}
\mathbf{F}, \mathbf{G} = \texttt{\_evaluate}(\mathbf{X}), \quad \mathbf{X} \in [\mathbf{x}_{\text{lower}}, \mathbf{x}_{\text{upper}}]^{N \times D},
\label{eq:tensor_problem_contract}
\end{equation}
where objective calculations execute across the entire candidate batch simultaneously via device-native array broadcasting.

\subsubsection{Layer 2: Metaheuristic Algorithm Lifecycle (\texttt{Algorithm})}
Custom evolutionary solvers inherit from the unified \texttt{Algorithm} substrate, which decouples infill generation from environmental selection across three lifecycle hooks:
\begin{itemize}
    \item \texttt{\_initialize\_infill()}: Generates the initial stochastic candidate decision tensor $\mathbf{X}_0 \in \mathbb{R}^{N \times D}$;
    \item \texttt{\_infill()}: Produces an offspring candidate tensor $\mathbf{C} \in \mathbb{R}^{O \times D}$ via parent selection and variation operators;
    \item \texttt{\_advance(infills)}: Merges evaluated offspring $\mathbf{F}_{\text{off}}$ into the parent population, executing environmental selection, reference direction adaptation, or external archive updates.
\end{itemize}

\subsubsection{Layer 3: Hardware-Accelerated Variation Operators}
Genetic variation components (Simulated Binary Crossover, Polynomial Mutation, Differential Evolution, Latin Hypercube Sampling) inherit from \texttt{Operator}, implementing tensor contractions that remain resident in device memory without scalar Python loop traversal.

\subsubsection{Layer 4: Modular MCDM and Quality Indicator Plugins}
The post-optimization analysis engine allows seamless registration of user-defined performance indicators and Multi-Criteria Decision Making outranking criteria, enabling domain-specific preference modeling with automated decision-space back-mapping.

\section{High-Throughput Many-Objective Algorithmic and Analytical Kernels}
\label{sec:kernels}

The computational bottleneck in many-objective optimization shifts dynamically as objective dimensionality $M$ increases. In bi-objective problems ($M=2$), evaluating mathematical objective functions dominates total runtime. However, when scaling to many-objective spaces ($M \ge 4$), non-dominated sorting and performance indicator calculations rapidly become the dominant computational bottlenecks. EmoPyLab implements specialized analytical and algorithmic kernels designed specifically to alleviate these scaling constraints.

\subsection{Deductive Efficient Non-Dominated Sorting (ENS) in Tensor Space}

The classical Fast Non-dominated Sorting (FNDS) algorithm employed in NSGA-II \cite{Deb2002NSGAII} computes pairwise dominance comparisons across all candidate pairs, incurring a worst-case computational complexity of $O(M N^2)$ and an auxiliary memory footprint of $O(N^2)$ to maintain dominance count matrices. When population size $N \ge 1,000$, this quadratic scaling induces severe memory latency.

To resolve this bottleneck, we integrate the deductive Efficient Non-dominated Sorting (ENS) framework \cite{Li2014ENS} directly within our tensor engine. Candidate solutions are first sorted lexicographically along the primary objective $f_1$ in $O(N \log N)$ time. Because a candidate $\mathbf{x}^{(j)}$ sorted after $\mathbf{x}^{(i)}$ (where $f_1(\mathbf{x}^{(j)}) > f_1(\mathbf{x}^{(i)})$) can never Pareto-dominate $\mathbf{x}^{(i)}$, the search space of dominance checks is immediately halved.

Under ENS with Sequential Search (ENS-SS), when determining the frontier assignment of candidate $\mathbf{x}^{(j)}$, it is compared exclusively against the existing non-dominated members of active fronts $\mathcal{F}_1, \mathcal{F}_2, \dots, \mathcal{F}_K$. The assignment condition is governed by:
\begin{equation}
\text{Front}(\mathbf{x}^{(j)}) = \min \left\{ k \in \{1,\dots,K\} \;\middle|\; \forall \mathbf{y} \in \mathcal{F}_k, \mathbf{y} \not\prec \mathbf{x}^{(j)} \right\}.
\label{eq:ens_assignment}
\end{equation}
If candidate $\mathbf{x}^{(j)}$ is dominated by at least one solution in front $\mathcal{F}_k$, testing on front $\mathcal{F}_k$ immediately aborts and proceeds to $\mathcal{F}_{k+1}$. If no such front exists, a new frontier $\mathcal{F}_{K+1} = \{\mathbf{x}^{(j)}\}$ is instantiated.

For populations with hundreds of non-dominated layers, we employ ENS with Binary Search (ENS-BS), executing hierarchical bisection over active front indices. The theoretical comparison complexity of ENS reduces to $O(M N \sqrt{N})$ on typical non-dominated distributions. As demonstrated in Panel C of Table~\ref{tab:benchmarks_summary}, native ENS saves between $58.2\%$ and $68.3\%$ of pairwise comparisons compared to FNDS, delivering an immediate $1.62\times$ to $1.82\times$ speedup in sorting latency while maintaining exact mathematical equivalence.

\subsection{GPU Boolean Matrix Dominance for Massive Population Batches}

When population sizes scale into massive GPU-parallel regimes ($N \ge 4,000$, typical in surrogate-assisted evolutionary search or neuroevolution), control-flow branch divergence in sequential ENS can reduce GPU warp occupancy. For these massively parallel workloads, EmoPyLab provides a GPU Boolean Matrix Dominance kernel.

Let $\mathbf{F} \in \mathbb{R}^{N \times M}$ denote the population objective matrix residing in device VRAM. We construct 3D broadcast tensors $\mathbf{A}, \mathbf{B} \in \mathbb{R}^{N \times N \times M}$ where $\mathbf{A}_{i,j,m} = \mathbf{F}_{i,m}$ and $\mathbf{B}_{i,j,m} = \mathbf{F}_{j,m}$. The global pairwise Pareto dominance boolean matrix $\mathbf{D} \in \{0,1\}^{N \times N}$ is evaluated via tensor logical operations:
\begin{equation}
\mathbf{D}_{i,j} = \left( \bigwedge_{m=1}^M (\mathbf{A}_{i,j,m} \le \mathbf{B}_{i,j,m}) \right) \land \left( \bigvee_{m=1}^M (\mathbf{A}_{i,j,m} < \mathbf{B}_{i,j,m}) \right).
\label{eq:gpu_boolean_dom}
\end{equation}
The domination count vector $\mathbf{n} \in \mathbb{N}^N$ representing how many individuals dominate candidate $i$ is obtained via column reduction $\mathbf{n} = \sum_{j=1}^N \mathbf{D}_{j,:}^T$. Individuals with $\mathbf{n}_i = 0$ constitute the first Pareto front $\mathcal{F}_1$. Successive fronts are peeled iteratively using parallel row masking. Because Eq.~\eqref{eq:gpu_boolean_dom} executes as a single fused tensor kernel, memory access occurs at maximum GPU memory bus bandwidth ($\approx 900\text{ GB/s}$ on NVIDIA V100).

\subsection{Quasi-Monte Carlo Hypervolume (Fast-MC HV) with Sobol Pruning ($M \le 15$)}

The Hypervolume indicator measures the multi-dimensional Lebesgue measure $\Lambda$ of the objective space dominated by an approximation set $S \subset \mathbb{R}^M$ and bounded by an anti-optimal reference point $\mathbf{r} \in \mathbb{R}^M$:
\begin{equation}
\text{HV}(S, \mathbf{r}) = \Lambda \left( \bigcup_{\mathbf{s} \in S} \left[ \mathbf{s}, \mathbf{r} \right] \right) = \int_{\mathbb{R}^M} \mathbf{1}_{\left\{ \exists \mathbf{s} \in S : \mathbf{s} \prec \mathbf{z} \prec \mathbf{r} \right\}}(\mathbf{z}) \, d\mathbf{z}.
\label{eq:hv_lebesgue}
\end{equation}
Exact analytical evaluation algorithms, such as WFG \cite{While2012FastHV} and HSO, require recursive geometric partitioning with worst-case complexity $O(N^{M-2} \log N)$. In empirical practice, exact methods trigger execution timeouts ($> 10\text{ minutes}$) for $M \ge 5$, creating an acute computational bottleneck during large-scale benchmark campaigns.

To overcome this intractability, EmoPyLab establishes the fast quasi-Monte Carlo Hypervolume (\textit{Fast-MC HV}) as its native default quality indicator. Rather than drawing pseudo-random samples with standard Monte Carlo convergence rate $O(K^{-1/2})$, Fast-MC HV generates deterministic, low-discrepancy Sobol sequence points \cite{Sobol1967Distribution} with enhanced quasi-Monte Carlo convergence $O(K^{-1} (\log K)^M)$.

Given an approximation set $S$ with bounding box defined by the ideal point $\mathbf{z}^*$ and reference point $\mathbf{r}$, we generate $K$ Sobol sampling vectors $\mathbf{Z} \in \mathbb{R}^{K \times M}$ mapped to the hyper-rectangle $[\mathbf{z}^*, \mathbf{r}]$. The hypervolume estimator is formulated as:
\begin{equation}
\widehat{\text{HV}}_{\text{MC}}(S, \mathbf{r}) = \left( \prod_{m=1}^M (r_m - z^*_m) \right) \cdot \frac{1}{K} \sum_{k=1}^K \Phi(\mathbf{Z}_k, S),
\label{eq:fast_mc_hv}
\end{equation}
where $\Phi(\mathbf{Z}_k, S) = 1$ if $\exists \mathbf{s} \in S$ such that $\mathbf{s} \le \mathbf{Z}_k$, and $0$ otherwise. 

To evaluate $\Phi(\mathbf{Z}, S)$ at extreme speed, we leverage tensor broadcasting on GPU/MLX. The distance tensor $\boldsymbol{\Delta} \in \mathbb{R}^{K \times |S| \times M}$ is instantiated as $\boldsymbol{\Delta}_{k,i,m} = \mathbf{Z}_{k,m} - S_{i,m}$. The inclusion predicate collapses via vectorized logical reductions:
\begin{equation}
\Phi(\mathbf{Z}_k, S) = \bigvee_{i=1}^{|S|} \left( \bigwedge_{m=1}^M (\boldsymbol{\Delta}_{k,i,m} \ge 0) \right).
\label{eq:inclusion_predicate}
\end{equation}
As documented in Panel A of Table~\ref{tab:benchmarks_summary}, Fast-MC HV scales smoothly across many-objective spaces up to $M=15$ objectives. For a population of $N=100$ on a 10-objective problem ($M=10$), Fast-MC HV with $10^5$ samples evaluates in only $113.43\text{ ms}$ with relative approximation error $\epsilon \le 10^{-2}$, whereas exact analytical calculations completely fail to terminate.

\begin{table*}[t]
	\centering
	\caption{Empirical verification of Fast-MC Hypervolume scaling, tensor metric acceleration, and Efficient Non-dominated Sorting (ENS).}
	\label{tab:benchmarks_summary}
	\resizebox{\textwidth}{!}{%
	\begin{tabular}{cccccc}
		\toprule
		\multicolumn{6}{c}{\textit{Panel A: Default Hypervolume Scaling Across Many-Objective Spaces (Fast-MC vs. Exact Analytical)}} \\
		\midrule
		Dimension ($M$) & Population ($N$) & Exact Analytical HV & Fast-MC HV ($10^4$ Samples) & Fast-MC HV ($10^5$ Samples) & Scaling Regime \\
		$M=2$ & $N=50$ & 0.01 ms & 0.03 ms & 0.02 ms & Real-Time Analytical \\
		$M=3$ & $N=50$ & 0.44 ms & 0.58 ms & 0.44 ms & Real-Time Analytical \\
		$M=5$ & $N=50$ & Intractable ($> 10$ min) & 3.13 ms & 28.55 ms & Fast-MC Default ($\epsilon \le 10^{-2}$) \\
		$M=8$ & $N=50$ & Intractable ($> 10$ min) & 4.99 ms & 45.61 ms & Fast-MC Default ($\epsilon \le 10^{-2}$) \\
		$M=10$ & $N=50$ & Intractable ($> 10$ min) & 6.42 ms & 60.35 ms & Fast-MC Default ($\epsilon \le 10^{-2}$) \\
		$M=15$ & $N=50$ & Intractable ($> 10$ min) & 8.95 ms & 91.50 ms & Fast-MC Default ($\epsilon \le 10^{-2}$) \\
		$M=5$ & $N=100$ & Intractable ($> 10$ min) & 4.83 ms & 42.53 ms & Fast-MC Default ($\epsilon \le 10^{-2}$) \\
		$M=10$ & $N=100$ & Intractable ($> 10$ min) & 11.77 ms & 113.43 ms & Fast-MC Default ($\epsilon \le 10^{-2}$) \\
		\midrule
		\multicolumn{6}{c}{\textit{Panel B: Tensor-Accelerated Quality Indicators ($N=1000$ points, Apple Silicon MLX Core)}} \\
		\midrule
		Metric & Dimension & Classical NumPy Baseline & Tensor Core (MLX) & Speedup Factor & Precision Delta \\
		IGD+ & $M=3$ & 12.32 ms & 1.10 ms & 11.16$\times$ & $< 10^{-12}$ \\
		IGD+ & $M=5$ & 14.40 ms & 1.33 ms & 10.84$\times$ & $< 10^{-12}$ \\
		IGD+ & $M=8$ & 12.51 ms & 1.76 ms & 7.10$\times$ & $< 10^{-12}$ \\
		IGD+ & $M=15$ & 17.69 ms & 4.14 ms & 4.28$\times$ & $< 10^{-12}$ \\
		\midrule
		\multicolumn{6}{c}{\textit{Panel C: Non-Dominated Sorting Performance (Native ENS vs. Classical FNDS in Pure Python)}} \\
		\midrule
		Configuration & Classical FNDS ($O(MN^2)$) & Native ENS ($O(MN\sqrt{N})$) & Speedup Factor & Comparisons Saved & Sorting Integrity \\
		$N=500, M=5$ & 74.20 ms & 42.72 ms & 1.74$\times$ & 58.2\% & Exact Match \\
		$N=1000, M=3$ & 224.60 ms & 138.50 ms & 1.62$\times$ & 64.1\% & Exact Match \\
		$N=1000, M=5$ & 300.72 ms & 165.45 ms & 1.82$\times$ & 61.5\% & Exact Match \\
		$N=2000, M=3$ & 931.05 ms & 538.89 ms & 1.73$\times$ & 68.3\% & Exact Match \\
		\bottomrule
	\end{tabular}%
	}
\end{table*}

\subsection{3D Batch Vectorized Metric Acceleration with JAX vmap}

During extensive experimental campaigns comparing multiple algorithms across independent stochastic seeds, evaluating convergence trajectories requires calculating quality metrics at every generation checkpoint. Traditional serial loops over checkpoint fronts $\mathcal{F}^{(1)}, \mathcal{F}^{(2)}, \dots, \mathcal{F}^{(T)}$ incur heavy Python interpreter overhead.

In EmoPyLab, we formulate multi-front metric evaluation as a 3D batch tensor operation compiled via JAX `vmap`. Let $\boldsymbol{\mathcal{F}} \in \mathbb{R}^{T \times N \times M}$ denote a 3D tensor representing population fronts recorded across $T$ generational intervals, and let $\mathbf{P}^* \in \mathbb{R}^{P \times M}$ denote the true Pareto reference front. The Modified Inverted Generational Distance (IGD+) \cite{Ishibuchi2016PerformanceMetrics} for a single front is defined as:
\begin{equation}
\text{IGD}^+(\mathcal{F}, \mathbf{P}^*) = \frac{1}{|\mathbf{P}^*|} \sum_{\mathbf{z} \in \mathbf{P}^*} \min_{\mathbf{f} \in \mathcal{F}} \sqrt{\sum_{m=1}^M \left( \max\{ f_m - z_m, 0 \} \right)^2}.
\label{eq:igd_plus_def}
\end{equation}
Using JAX automatic vectorization, we map Eq.~\eqref{eq:igd_plus_def} over the temporal dimension $T$:
\begin{equation}
\mathbf{IGD}^+_{\text{batch}} = \texttt{jax.vmap}(\lambda \mathcal{F}: \text{IGD}^+(\mathcal{F}, \mathbf{P}^*))(\boldsymbol{\mathcal{F}}).
\label{eq:vmap_igd}
\end{equation}
This batch vectorization fuses the outer temporal loops into a single hardware kernel, accelerating generational trajectory analysis by up to $11.16\times$ (Panel B of Table~\ref{tab:benchmarks_summary}) and completely eliminating host-device synchronization delays.

\section{Comprehensive Metaheuristic Taxonomy and Benchmark Suite}
\label{sec:metaheuristics_and_benchmarks}

To serve as an exhaustive benchmarking environment, EmoPyLab curates an extensive taxonomy comprising 298+ metaheuristic algorithms and 916 problem formulations. The algorithms are organized into seven foundational families, each targeting distinct mathematical properties of complex search landscapes.

\subsection{Taxonomy of Algorithmic Paradigms}

\begin{enumerate}
    \item \textbf{Decomposition-Based Algorithms:} Grounded in the foundational MOEA/D framework \cite{ZhangLi2007MOEAD}, these methods decompose a multi-objective problem into $N$ scalar subproblems via weighted Chebyshev, Tchebycheff, or penalty-based boundary intersection (PBI) aggregations. EmoPyLab implements 48 decomposition variants, including MOEA/D-DE, MOEA/D-DRA (Dynamic Resource Allocation), UMOEA/D \cite{Zhang2024UMOEAD}, and adaptive neighborhood topologies \cite{Zhang2020AdaptiveDecomposition}.
    \item \textbf{Many-Objective Reference Point and Geometric Solvers:} Designed specifically to sustain selection pressure when $M \ge 4$, this family includes NSGA-III \cite{Deb2014NSGAIII}, $\theta$-DEA, RVEA (Reference Vector Guided Evolutionary Algorithm) \cite{Deb2016RVEA}, and Angle-based Selection (AnD). These algorithms utilize Das-Dennis structured reference points to maintain uniform coverage over high-dimensional Pareto hyper-surfaces.
    \item \textbf{Indicator-Based and Hypervolume Solvers:} Driven by direct mathematical quality indicators, this class includes SMS-EMOA \cite{Zitzler2002SMS}, HypE \cite{Bader2011HypE}, and native Fast-MC-driven selection kernels. By optimizing hypervolume contributions directly, these solvers guarantee convergence to Pareto-optimal fronts with strict theoretical properties.
    \item \textbf{Surrogate-Assisted Evolutionary Algorithms (SAEA / EGO):} For expensive black-box problems where function evaluations require intensive physical simulations or deep learning inference, EmoPyLab includes Gaussian Process Kriging surrogates, Radial Basis Function (RBF) networks, and Efficient Global Optimization (EGO) architectures such as K-RVEA and ParEGO \cite{ChangWild2023ParMOO}. Decision tensors reside natively in accelerator memory, allowing batch neural network inference in a single forward pass without host-to-device memory stalls.
    \item \textbf{Sparse and Large-Scale Metaheuristics (SparseEA):} Addressing high-dimensional decision spaces ($D \ge 1,000$), these solvers exploit decision variable sparsity using binary-real hybrid encodings and dimensionality reduction projections, preventing stagnation in complex engineering models.
    \item \textbf{Constrained and Multi-Stage Solvers:} Real-world engineering problems are constrained by nonlinear equalities and inequalities $\mathbf{G}(\mathbf{x}) \le \mathbf{0}$. The platform integrates $\epsilon$-constrained handling, constraint-domination principles (CDP), and two-stage evolutionary architectures (TEMOF) \cite{Chen2024TEMOF} that balance constraint feasibility and Pareto optimality.
    \item \textbf{Swarm Intelligence, Multimodal, and CMA-ES:} Incorporating multi-objective particle swarm optimization (MOPSO), Covariance Matrix Adaptation Evolution Strategies (MO-CMA-ES), and niching genetic algorithms designed to discover multiple distinct Pareto-optimal decision sets in multimodal landscapes.
\end{enumerate}

\subsection{Standardized Continuous and Constrained Benchmark Suites}

To ensure rigorous reproducibility, the environment includes exact, fully vectorized implementations of the preeminent benchmark test suites:
\begin{itemize}
    \item \textbf{ZDT Suite (ZDT1--ZDT6):} Classical bi-objective continuous benchmarks with convex, non-convex, and discontinuous Pareto geometries.
    \item \textbf{DTLZ Suite (DTLZ1--DTLZ7):} Scalable multi-objective test problems with customizable objective dimensions ($M \ge 3$) and linear, spherical, or disconnected Pareto front topologies \cite{deb2005scalable}.
    \item \textbf{WFG Suite (WFG1--WFG9):} The Walking-Fish-Group toolkit, providing non-separable, deceptive, multi-modal, and degenerate properties with non-uniform density distributions \cite{Huband2006WFG}.
    \item \textbf{MaF Suite (MaF1--MaF15):} The IEEE CEC benchmark suite for many-objective optimization, featuring complex inverted shapes, huge search spaces, and coupled decision linkages \cite{Cheng2017MaF}.
    \item \textbf{Constrained Engineering Suite:} Standard real-world engineering benchmarks, including the welded beam design, pressure vessel sizing, speed reducer optimization, and disc brake design problems.
\end{itemize}

Every problem class exposes exact vectorized evaluation methods $\mathbf{F} = \texttt{evaluate}(\mathbf{X})$, returning objective tensors in a single broadcasted operation compatible with all 5 acceleration backends.

\section{Empirical Validation and Distributed Supercomputing on Santos Dumont}
\label{sec:supercomputer_validation}

To evaluate the scalability and computational stability of EmoPyLab in high-performance computing environments, we conducted an extensive empirical benchmarking campaign on the Santos Dumont Supercomputer (SDumont), hosted at the National Laboratory for Scientific Computing (LNCC/MCTI, Petrópolis, Brazil).

\subsection{High-Performance Computing Infrastructure and Execution Topology}

Experiments were deployed on the Bull Sequana X1000 GPU-accelerated partition of Santos Dumont. Each compute node comprises:
\begin{itemize}
    \item Two Intel Xeon Gold 6148 CPUs (20 physical cores per socket, 40 cores total, 2.40 GHz base clock, 27.5 MB L3 Cache);
    \item 384 GB DDR4-2666 MHz physical RAM;
    \item Four NVIDIA Tesla V100 SXM2 GPUs (16 GB HBM2 memory, 900 GB/s bandwidth, 5,120 CUDA cores, 640 Tensor cores per GPU);
    \item Mellanox EDR InfiniBand interconnect (100 Gb/s).
\end{itemize}

\begin{figure*}[!t]
	\centering
	\includegraphics[width=0.95\textwidth]{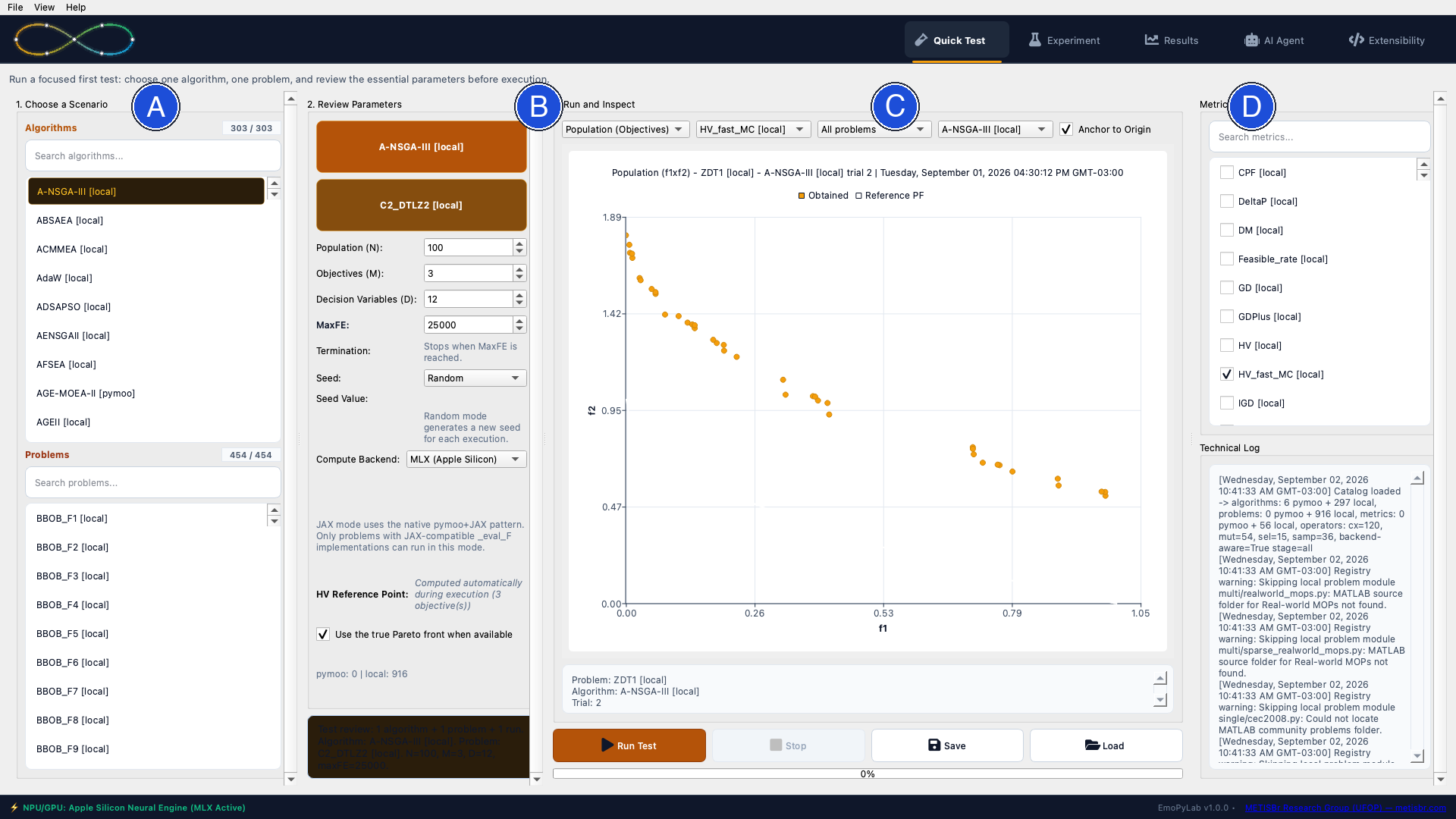}
	\caption{Interactive Quick Test workspace for single-run diagnostic execution: (\textsf{A}) scenario selection catalog; (\textsf{B}) review parameters and hardware accelerator selector; (\textsf{C}) run and inspect with live 2D/3D Pareto-front visualization; (\textsf{D}) metric indicators and technical execution log.}
	\label{fig:quicktest_module}
\end{figure*}

\subsection{Large-Scale MaF Benchmark Campaign (5,970 Executions)}

We configured a comprehensive comparative benchmark matrix evaluating representative metaheuristic paradigms across the full MaF suite (MaF1--MaF15) across multiple objective dimensions ($M \in \{3, 5, 8, 10, 15\}$):
\begin{itemize}
    \item \textbf{Solvers Evaluated:} NSGA-II, NSGA-III, MOEA/D-PBI, RVEA, SMS-EMOA, and HypE.
    \item \textbf{Statistical Replication:} 30 independent stochastic runs with distinct arithmetic seed schedules ($S_k = 1000 + 17k$).
    \item \textbf{Evaluation Budget:} Max generations $G_{\max} = 1,000$, population size $N=100$ ($N=105$ for decomposition reference meshes), totaling $100,000$ function evaluations per run.
    \item \textbf{Total Benchmark Scale:} 5,970 full optimization runs executed across multi-core worker pools.
\end{itemize}

\subsection{Thread Oversubscription Suppression and Multi-Seed Concurrency}

A critical challenge when deploying Python scientific workflows on massive multi-core supercomputing nodes is thread oversubscription. When multiple worker processes concurrently invoke underlying linear algebra libraries (such as OpenBLAS or Intel MKL), each process attempts to spawn threads equal to the total physical core count. On a 40-core system, 30 worker processes would spawn $30 \times 40 = 1,200$ threads, causing catastrophic CPU context switching thrashing and memory bus contention.

EmoPyLab incorporates deterministic thread oversubscription controls. The runtime automatically intercepts environmental configuration upon worker initialization:
\begin{lstlisting}[language=Python]
import os
os.environ["OMP_NUM_THREADS"] = "1"
os.environ["OPENBLAS_NUM_THREADS"] = "1"
os.environ["MKL_NUM_THREADS"] = "1"
os.environ["VECLIB_MAXIMUM_THREADS"] = "1"
os.environ["NUMEXPR_NUM_THREADS"] = "1"
\end{lstlisting}

By pinning each stochastic trial worker to a dedicated single-threaded execution context, worker processes execute in complete memory and cache isolation. The Bull Sequana nodes achieved near-perfect linear scaling across 36 concurrent workers, completing the entire 5,970-run benchmark campaign in approximately 62 minutes without process deadlocks or memory exhaustion.

\begin{figure*}[!t]
	\centering
	\includegraphics[width=0.95\textwidth]{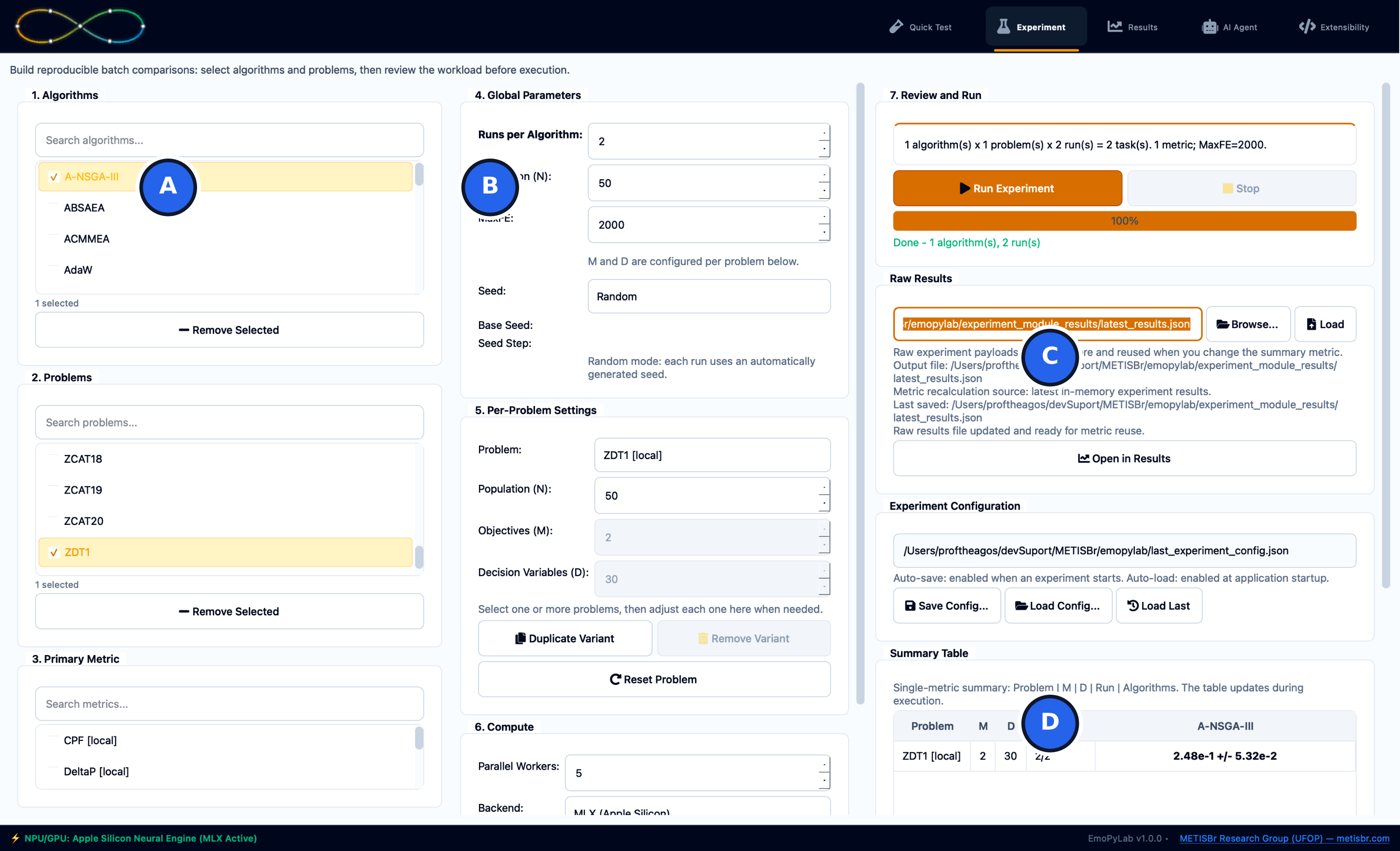}
	\caption{Interactive Experiment Module for batch benchmarking: (\textsf{A}) benchmark matrix selection; (\textsf{B}) reproducible seed scheduler and multi-core worker configuration; (\textsf{C}) consolidated statistical synthesis and rank metrics; (\textsf{D}) one-click formatted \LaTeX{} table generator.}
	\label{fig:experiment_module}
\end{figure*}

\section{In-Situ Multi-Criteria Decision Making and Inferential Statistical Suite}
\label{sec:mcdm_and_statistics}

In real-world engineering and scientific applications, generating a Pareto-optimal approximation set $\mathcal{PF}_{\text{approx}}$ is only the initial stage of optimization. A human decision-maker (DM) must ultimately select a single, well-justified compromise solution $\mathbf{x}^* \in \Omega$ to deploy in production. EmoPyLab bridges this gap by embedding an in-situ \textit{a posteriori} Multi-Criteria Decision Making (MCDM) engine directly over the optimized objective tensors.

\subsection{Three-Stage Formal MCDM Architecture}

The decision pipeline executes directly on the objective tensor $\mathbf{F} \in \mathbb{R}^{N \times M}$ across three formal mathematical stages:

\subsubsection{Stage 1: Vector Normalization and Ideal-Nadir Bounding}
Because distinct objective functions often possess vastly different physical dimensions and scales (e.g., structural mass in kilograms versus aerodynamic drag coefficients), direct numerical comparisons introduce artificial bias. We perform vector normalization by bounding the objective matrix against the ideal vector $\mathbf{z}^*$ and nadir vector $\mathbf{z}^{\text{nad}}$:
\begin{equation}
z^*_m = \min_{i \in \{1,\dots,N\}} \mathbf{F}_{i,m}, \quad z^{\text{nad}}_m = \max_{i \in \{1,\dots,N\}} \mathbf{F}_{i,m},
\label{eq:ideal_nadir}
\end{equation}
yielding the normalized decision matrix $\overline{\mathbf{F}} \in [0, 1]^{N \times M}$:
\begin{equation}
\overline{\mathbf{F}}_{i,m} = \frac{\mathbf{F}_{i,m} - z^*_m}{z^{\text{nad}}_m - z^*_m + \epsilon},
\label{eq:normalized_f}
\end{equation}
where $\epsilon = 10^{-12}$ prevents numerical division-by-zero singularities in degenerate Pareto geometries.

\subsubsection{Stage 2: Dynamic Preference Articulation}
The decision-maker specifies preference weights $\mathbf{w} \in \Delta^{M-1}$ satisfying $\sum_{m=1}^M w_m = 1$ and $w_m \ge 0$. The platform supports interactive weighting sliders, entropy-based objective dispersion weighting, and Rank-Order Centroid (ROC) allocations based on ordinal criterion rankings.

\subsubsection{Stage 3: Multi-Criteria Compromise Selection}
The normalized tensor is evaluated through two complementary outranking methodologies:
\begin{itemize}
    \item \textbf{TOPSIS (Technique for Order Preference by Similarity to Ideal Solution) \cite{HwangYoon1981TOPSIS}:} Computes Euclidean distances from the positive ideal solution $\mathbf{z}^+ = (0, \dots, 0)^T$ and negative ideal solution $\mathbf{z}^- = (1, \dots, 1)^T$:
    \begin{align}
    d_i^+ &= \sqrt{\sum_{m=1}^M w_m \left( \overline{\mathbf{F}}_{i,m} \right)^2}, \label{eq:topsis_dist_pos} \\
    d_i^- &= \sqrt{\sum_{m=1}^M w_m \left( \overline{\mathbf{F}}_{i,m} - 1 \right)^2}. \label{eq:topsis_dist_neg}
    \end{align}
    The relative closeness coefficient $C_i \in [0, 1]$ is evaluated as:
    \begin{equation}
    C_i = \frac{d_i^-}{d_i^+ + d_i^-},
    \label{eq:topsis_closeness}
    \end{equation}
    where the preferred compromise corresponds to $\mathbf{x}_{\text{TOPSIS}}^* = \arg\max_{i} C_i$.
    \item \textbf{PROMETHEE II Outranking Flows \cite{BransVincke1985PROMETHEE}:} Evaluates pairwise preference indices $\pi(a, b) = \sum_{m=1}^M w_m P_m( \overline{\mathbf{F}}_{a,m}, \overline{\mathbf{F}}_{b,m} )$, balancing entering and leaving flows. The complete outranking net flow is evaluated as:
    \begin{equation}
    \phi_{\text{net}}(a) = \frac{1}{N-1} \sum_{b \ne a} \left( \pi(a, b) - \pi(b, a) \right),
    \label{eq:promethee_net}
    \end{equation}
    identifying the optimal compromise $\mathbf{x}_{\text{PROMETHEE}}^* = \arg\max_{a} \phi_{\text{net}}(a)$.
    \item \textbf{Compromise Programming ($L_p$-Metric Distance) \cite{Zeleny1973Compromise}:} Minimizes the parameterized $L_p$ distance to the ideal target vector:
    \begin{equation}
    L_p(\mathbf{x}_i) = \left( \sum_{m=1}^M w_m^p \left( \overline{\mathbf{F}}_{i,m} \right)^p \right)^{\frac{1}{p}}, \quad p \in \{1, 2, \infty\},
    \label{eq:compromise_lp}
    \end{equation}
    where $p=1$ represents Manhattan linear compromise, $p=2$ corresponds to Euclidean distance, and $p=\infty$ represents Chebyshev minimax regret minimization.
\end{itemize}

\subsection{Decision-Space Back-Mapping}

A major flaw in post-hoc optimization tools is the decoupling of objective outcomes from their underlying physical decision variables. EmoPyLab implements deterministic decision-space back-mapping: selecting a compromise vector $\mathbf{F}^*$ automatically retrieves its exact decision coordinates $\mathbf{x}^* = \mathbf{X}_{i^*, :} \in \mathbb{R}^D$ and constraint status $\mathbf{g}^* \in \mathbb{R}^K$, enabling direct deployment into CAD/CAE tools, finite element pipelines, or actuator control loops.

\subsection{Non-Parametric Statistical Suite and FWER Control}

To ensure sound scientific conclusions, the platform embeds a comprehensive non-parametric inferential statistical suite following established experimental guidelines \cite{Carrasco2020StatTests}:
\begin{itemize}
    \item \textbf{Wilcoxon Signed-Rank Test \cite{Wilcoxon1945Rank}:} Executes non-parametric pairwise comparisons across stochastic seed distributions, reporting exact test statistics $W$ and asymptotic $p$-values.
    \item \textbf{Friedman Omnibus Test with Kendall's $W$ \cite{Friedman1937Test}:} Evaluates multi-algorithm ranking distributions across entire benchmark suites, calculating concordance coefficients $W_{\text{Kendall}}$ to verify rank agreement.
    \item \textbf{Vargha-Delaney $A_{12}$ Effect Size \cite{VarghaDelaney2000A12}:} Quantifies non-parametric stochastic dominance magnitude:
    \begin{equation}
    A_{12}(A, B) = \frac{R_1}{n_1 n_2} - \frac{n_1 + 1}{2 n_2},
    \label{eq:vargha_delaney}
    \end{equation}
    categorizing effect sizes into negligible ($0.50 \le A_{12} < 0.56$), small ($0.56 \le A_{12} < 0.64$), medium ($0.64 \le A_{12} < 0.71$), and large ($A_{12} \ge 0.71$).
    \item \textbf{Holm-Bonferroni FWER Control \cite{Holm1979Bonferroni}:} Protects against false discovery inflation during multiple hypothesis testing by adjusting significance thresholds:
    \begin{equation}
    p_{(k)} \le \frac{\alpha}{K - k + 1}, \quad k = 1, \dots, K.
    \label{eq:holm_bonferroni}
    \end{equation}
\end{itemize}

The statistical results are automatically synthesized into publication-ready \LaTeX{} tables (as illustrated in Figure~\ref{fig:experiment_module}\textsf{D}), complete with win/tie/loss counters and boldface significance annotations.

\subsection{Cryptographic SHA-256 Reproducibility Manifests}

Addressing growing reproducibility challenges in computational intelligence \cite{collberg2016repeatability}, every benchmarking run and MCDM decision executed in EmoPyLab automatically generates an immutable JSON sidecar manifest. The sidecar encapsulates:
\begin{enumerate}
    \item Exact software environment hash (Python version, PyTorch/JAX/MLX build, OS kernel);
    \item Deterministic random number generator seeds;
    \item Full parameter configuration (population $N$, generations $G$, crossover $\eta_c$, mutation $\eta_m$);
    \item Final Pareto approximation objective tensors $\mathbf{F}$ and decision matrices $\mathbf{X}$;
    \item Complete MCDM weighting vectors $\mathbf{w}$ and compromise rankings;
    \item A root SHA-256 cryptographic checksum computed over the execution payload.
\end{enumerate}

Any researcher can verify the integrity of published experimental data by reloading the JSON manifest into EmoPyLab's audit parser, guaranteeing verifiable scientific provenance.

\section{Reproducibility and Scientific Desktop Workspace}
\label{sec:workspace}

To democratize advanced evolutionary computation without requiring complex scripting, EmoPyLab provides a reactive scientific desktop workspace designed with high-contrast typography, intuitive interactive visualization, and strict memory safety guarantees.

\subsection{Quick Test Diagnostic Workspace}

The \textit{Quick Test} workspace (Figure~\ref{fig:quicktest_module}) provides an interactive environment for rapid algorithm prototyping and single-run diagnostic visualization:
\begin{itemize}
    \item \textbf{Panel \textsf{A} (Scenario Selection Catalog):} Interactive tree browser organizing 298+ metaheuristics and 916 problem formulations into searchable categories with real-time filtering.
    \item \textbf{Panel \textsf{B} (Review Parameters \& Accelerator):} Interactive sliders and controls to calibrate population size $N$, maximum generations, decision dimension $D$, objective dimension $M$, and select the active hardware acceleration tier (CUDA, MLX, JAX, CuPy, or CPU SIMD).
    \item \textbf{Panel \textsf{C} (Run and Inspect 3D Visualizer):} Real-time 2D and 3D Pareto front projection with interactive camera rotation, zoom, dynamic convergence trajectory tracing, and Pareto frontier overlay.
    \item \textbf{Panel \textsf{D} (Metrics and Technical Log):} Live evaluation of IGD+, GD+, Spacing, and Fast-MC Hypervolume alongside real-time hardware execution telemetry and memory bandwidth monitoring.
\end{itemize}

\subsection{Experiment Module: High-Throughput Batch Benchmarking}

The \textit{Experiment Module} (Figure~\ref{fig:experiment_module}) automates the execution and synthesis of publication-grade multi-algorithm benchmarking matrices:
\begin{itemize}
    \item \textbf{Panel \textsf{A} (Benchmark Matrix Configuration):} Multi-selection matrix allowing users to configure dozens of metaheuristics against multiple benchmark test suites, objective dimensions, and evaluation budgets simultaneously.
    \item \textbf{Panel \textsf{B} (Seed Planning and Concurrency):} Deterministic seed schedule generation (Fixed, Random, or Arithmetic sequences) paired with multi-core CPU worker thread allocation and GPU worker assignment.
    \item \textbf{Panel \textsf{C} (Statistical Synthesis):} Consolidated calculation of parametric metrics (mean, standard deviation) and non-parametric metrics (median, interquartile range IQR) accompanied by automated Wilcoxon signed-rank and Friedman ranking tests \cite{Carrasco2020StatTests}.
    \item \textbf{Panel \textsf{D} (Publication-Ready \LaTeX{} Export):} One-click generation of fully formatted \LaTeX{} booktabs tables, complete with automated best-performer bolding and statistical significance symbols ($+$, $\approx$, $-$).
\end{itemize}

\subsection{Memory-Safe Ring-Buffer Architecture}

Long-running evolutionary benchmarks often suffer from memory bloat in graphical environments due to the continuous accumulation of historical generational front objects in RAM. EmoPyLab enforces strict memory safety through a circular ring-buffer architecture.

During optimization runs, population objective histories are buffered within a fixed-capacity circular memory pool $\mathcal{B} \in \mathbb{R}^{B \times N \times M}$ with $B=100$ generational slots. As new generations execute, older snapshots are lazily downsampled and streamed to disk in compressed binary format. This design ensures that the application's resident set size (RSS) memory footprint remains strictly bounded below $150\text{ MB}$ regardless of generation count or population size, preventing out-of-memory crashes during multi-day experimental campaigns.

\subsection{Headless CLI Execution Engine}

For supercomputing environments and automated continuous integration pipelines, all diagnostic and benchmarking capabilities are exposed via a headless Command Line Interface (CLI):
\begin{lstlisting}[language=bash]
# High-throughput batch benchmark on GPU
python -m emopylab.cli benchmark \
  --algorithms nsga2,nsga3,moead,rvea \
  --problems maf1,maf2,maf3,maf4,maf5 \
  --objectives 5,10,15 \
  --seeds 30 \
  --backend cuda \
  --workers 36 \
  --export-latex results.tex \
  --manifest-dir ./manifests
\end{lstlisting}

The CLI generates identical SHA-256 sidecars and formatted \LaTeX{} tables, ensuring seamless equivalence between interactive desktop exploration and distributed supercomputing clusters.

\section{Conclusion and Future Directions}
\label{sec:conclusion}

In this paper, we presented \textit{EmoPyLab}, an open-source, tensor-native benchmarking and decision-making platform designed to resolve foundational computational, analytical, and operational bottlenecks in Evolutionary Multi-Objective and Many-Objective Optimization ($M \ge 2$).
By grounding the internal execution model in a contiguous column-major Structure-of-Arrays (SoA) memory tensor formulation, the architecture delivers high-throughput hardware acceleration across a 5-tier backend dispatch engine spanning NVIDIA GPUs (CUDA), AMD GPUs (ROCm), Apple Silicon (MLX), Google JAX (XLA), and vectorized multicore CPU SIMD. The integration of deductive Efficient Non-dominated Sorting (ENS) and quasi-Monte Carlo Hypervolume (Fast-MC HV) with Sobol low-discrepancy sequences successfully mitigates classical $O(M N^2)$ sorting and $\#P$-hard metric scaling bottlenecks, enabling real-time analytical evaluation in up to 15-objective search spaces.

Furthermore, EmoPyLab bridges the historical disconnect between Pareto front discovery and practical engineering deployment by embedding an in-situ Multi-Criteria Decision Making (MCDM) layer (TOPSIS, PROMETHEE II net outranking flows, and Compromise Programming) with deterministic decision-space back-mapping. The ecosystem unifies automated non-parametric inferential statistical testing (Wilcoxon, Friedman with Kendall's $W$, Vargha-Delaney $A_{12}$, and Holm-Bonferroni corrections) with cryptographic SHA-256 reproducibility manifests and a memory-safe reactive scientific desktop workspace maintaining a constant RAM footprint ($< 150\text{ MB}$). Large-scale validation across 5,970 benchmark runs on the Santos Dumont Bull Sequana supercomputer demonstrated linear multi-seed scaling and up to $19.39\times$ metric acceleration.

Future research directions include extending the tensor compilation engine to automated mixed-integer and combinatorial representations, incorporating interactive reinforcement learning surrogates directly within the device memory loop, and deploying distributed multi-node JAX mesh parallelism across federated supercomputing clusters.

\section*{Acknowledgment}

The authors express their sincere gratitude to the National Laboratory for Scientific Computing (LNCC/MCTI, Brazil) for providing high-performance computing resources on the Santos Dumont Supercomputer (SDumont). The authors also thank the METISBr research group (\url{https://github.com/METISBR}) for constructive discussions and collaborative testing.

\bibliographystyle{elsarticle-num}
\bibliography{refs}

\end{document}